\documentstyle[aps,epsfig,manuscript,multicol]{revtex}
\begin{document}
\newcommand{\beq}{\begin{equation}}
\newcommand{\eeq}{\end{equation}}
\newcommand{\sg}{\sigma_p}
\newcommand{\sm}{\sigma_c}
\newcommand{\lf}{\langle L_f \rangle}
\newcommand{\la}{\langle L_a \rangle}
\newcommand{\tf}{{\rm Tr}_f}
\newcommand{\nt}{N_\tau}
\newcommand{\ns}{N_\sigma}
\newcommand{\f}{\beta_f}
\newcommand{\ba}{\beta_a}
\newcommand{\bv}{\beta_v}
\newcommand{\bt}{\beta}
\newcommand{\bvc}{\beta_{vc}}
\newcommand{\lm}{\lambda}

\title{Peak Effect in Ca$_3$Rh$_4$Sn$_{13}$ : Vortex Phase Diagram and Evidences for Stepwise Amorphization of Flux Line Lattice}

\author{S.~Sarkar$^1$\footnote{E-mail:shampa@tifr.res.in }, D.~Pal$^1$, S.~S.~Banerjee$^1$, 
S.~Ramakrishnan$^1$, A.~K.~Grover$^1$\footnote{E-mail:grover@tifr.res.in }, C.~V.~Tomy$^2$, G.~Ravikumar$^3$, P.~K.~Mishra$^3$, V.~C.~Sahni$^3$, G.~Balakrishnan$^4$, D.~McK.~Paul$^4$ and S.~Bhattacharya$^{1,5}$}

\address{$^1$Department of Condensed Matter Physics and Materials Science,\\ 
Tata Institute of Fundamental Research, Colaba, Mumbai 400005, India \\
$^2$ Department of Physics, Indian Institute of Technology, Powai, Mumbai  400076, India\\
$^3$Technical Physics and Prototype Engineering Division, Bhabha 
Atomic Research Centre, Mumbai 400 085, India \\
$^4$Department of Physics, University of Warwick, Coventry CV4 7AL, U.K.\\
$^5$NEC Research Institute, 4 Independence Way, Princeton, N. J. 08540}
\maketitle

\begin{abstract}

The Peak Effect (PE) regime in a single crystal of 
Ca$_3$Rh$_4$Sn$_{13}$ has been investigated in detail via 
ac susceptibility as well as dc magnetization 
measurements. The PE region comprises two 
discontinuous first order like transitions, 
located near its $onset$ and $peak$ positions and reflects a 
stepwise fracturing of the flux line 
lattice (FLL). This scenario 
can be construed as an alternative to a scheme involving 
the appearance of Generalized 
Fulde-Ferrel-Larkin-Ovchinnikov (GFFLO) 
state as considered in the case of PE 
phenomenon in isostructural Yb$_3$Rh$_4$Sn$_{13}$  
system. The 
procedure of thermal cycling across the onset 
position of PE produces an open 
hysteresis loop, which is consistent with the notion of 
the fracturing of the FLL. A study of thermomagnetic 
history dependence of different metastable vortex states shows 
the path dependence in the critical current density 
$J_c$$(H,T)$ over a large part of the $(H,T)$ parameter space. 
The path dependence in $J_c$ ceases above the $peak$ 
position of the PE, suggesting a complete amorphization 
and the loss of long range correlations of the FLL at ($T_p$, $H_p$) line. A 
plausible vortex
 phase diagram has been illustrated for Ca$_3$Rh$_4$Sn$_{13}$ and 
phases like an elastic solid, a plastic solid, a pinned amorphous and an unpinned amorphous states have been identified.
\end{abstract}
\clearpage
\section{INTRODUCTION}
\label{sec:INTRO}

Peak Effect (PE) phenomenon relates to an anomalous 
behavior of (macroscopic) current carrying capacity 
in the mixed state of a type II superconductor. 
 The critical current density $J_c$ of a superconductor, 
which usually decreases monotonically with increasing $H$ 
or $T$, can show an anomalous increase  before
proceeding to zero at or close to the superconducting to 
normal phase boundary ($H_{c2}$/$T_c(H)$ line). The 
occurrence of PE, ubiquitous in conventional low 
temperature superconductors, has received a considerable 
amount of attention in recent years 
\cite{r1,r2,r3,r4,r5,r6,r7,r8,r9,r10,r11,r12,r13,r14}. 
A large variety of 
superconducting 
systems exhibit PE, such as, Nb (pure metal) 
\cite{r7}, 
V$_3$Si ($A15$ compound) \cite{r8}, 2$H$-NbSe$_2$ (layered 
chalcogenide) \cite{r1,r10,r11}, 
rare earth intermetallics, like, CeCo$_2$\cite{r9}
 and CeRu$_2$\cite{r2,r5,r6}, Yb$_3$Rh$_4$Sn$_{13}$ 
(ternary stannide) \cite{r12,r13}, YNi$_2$B$_2$C 
(quarternary borocarbide) \cite{r14}, 
 UPd$_2$Al$_3$ (heavy fermion superconductor)\cite{r3,r4},  
YBa$_2$Cu$_3$O$_7$ 
(high temperature superconductor) \cite{r15,r16}, etc. Of the 
various scenarios being considered to understand PE 
phenomenon, two appear distinct and independent. A 
classical scenario, originally due to Pippard \cite{r17}, 
attempts to relate PE to a faster rate of decrease in 
elastic moduli of FLL with increase in temperature as 
compared to that of the elementary pinning force. A 
recent scenario due to Tachiki et al \cite{r4}, suggests 
that PE phenomenon in heavy fermion superconductors (HFS) 
could be caused as a first order transition by the 
realization of a generalized Fulde-Ferrel-Larkin-Ovchinnikov (GFFLO) \cite{r18,r19} state at high fields 
and low temperatures ($T$~$\le$~0.55$T_c$(0)). The rare 
earth and Uranium based superconductors have 
large normal state paramagnetism, and in the 
superconducting mixed state of these systems, the order 
parameter can assume a spatial modulation (with nodes in the 
longitudinal direction) at high fields such that the 
vortex line (then) comprises coupled segments of lengths 
comparable to wavelength of modulation in superconducting 
order parameter. The inevitably present pinning centres 
in the sample enhance their pinning action 
on such vortex lines at the onset of GFFLO transition and 
thereby cause PE phenomenon.\\

Modler et al \cite{r3} have shown that the 
characteristics of several physical properties in the PE 
region of a single crystal of the heavy fermion compound  
UPd$_2$Al$_3$ are similar to those in a single 
crystal of the mixed valent compound CeRu$_2$
 and the phase boundaries (locus of $H$ and $T$ 
values) drawn for UPd$_2$Al$_3$ and CeRu$_2$ extend over 
similar parametric limits in the $(H,T)$ space\cite{r20}. Sato et al 
\cite{r12}, followed by Tomy et al \cite{r13}, reported 
observation of the PE and the construction of a magnetic phase 
diagram in weakly pinned crystals of Yb$_3$Rh$_4$Sn$_{13}$,  
which bore striking resemblance to such phase diagrams in UPd$_2$Al$_3$ 
and CeRu$_2$\cite{r3}. On the other hand, Crabtree et al \cite{r21} 
have drawn attention to the similarities in the details 
of the transport studies on the PE phenomenon in single 
crystals of CeRu$_2$ and YBa$_2$Cu$_3$O$_7$ with 
those observed in crystals 
of 2$H$-NbSe$_2$ \cite{r1}. The latter system, 2$H$-NbSe$_2$, 
is thought to be an archetypal example for the classical scenario of collapse of 
elastic moduli of flux line lattice (FLL) (implying a thermal and/or disorder induced amorphization of  the FLL\cite{r1,r22}) as a source of PE. Very recently, 
Banerjee et al \cite{r23} have shown that in crystals of 
2$H$-NbSe$_2$ and CeRu$_2$, having comparable levels of 
pinning, the PE phenomenon comprises a rich structure that
can be construed as an evidence of a stepwise 
fracturing of the vortex lattice at or near the incipient FLL melting 
transition in both the systems. It is asserted \cite{r23} 
that the destruction of the 
elastically deformed vortex solid commences at the onset 
of PE via a pinning-induced fracturing transition and the 
amorphization of the vortex solid proceeds to completion 
at the peak position of PE due to juxtaposition of 
effects caused by thermal fluctuations and quenched 
random disorder. Between the onset and the peak positions 
of the PE, the vortex solid is believed to be in a 
plastically deformed state, as revealed by earlier 
transport studies in 2$H$-NbSe$_2$\cite{r1} and CeRu$_2$\cite{r21}.\\

Independent of these developments, Tomy et al \cite{r24} 
have reported the occurrence of PE and the construction of a 
magnetic phase diagram in Ca$_3$Rh$_4$Sn$_{13}$, which 
again bear striking resemblance to those in 
UPd$_2$Al$_3$ and CeRu$_2$\cite{r3}. This ternary system has the same crystal 
structure as the ternary rare earth stannide 
Yb$_3$Rh$_4$Sn$_{13}$\cite{r12,r13}, but, it does not contain any rare 
earth ions. Its normal state paramagnetic value of 
3.3$\times$10$^{-6}$~emu/cm$^3$ precludes the realization of a 
GFFLO state\cite{r4} in it. 
In view of an apparent similarity in the 
magnetic phase diagram of Ca$_3$Rh$_4$Sn$_{13}$ with those in  
other compounds like Yb$_3$Rh$_4$Sn$_{13}$, 2$H$-NbSe$_2$, 
CeRu$_2$, CeCo$_2$, UPd$_2$Al$_3$, etc., it is of 
interest to investigate the detailed behaviour in the PE region of 
Ca$_3$Rh$_4$Sn$_{13}$, and to compare them with the results of Banerjee et al
\cite{r23} and Ravikumar 
et al\cite{r25} in 2$H$-NbSe$_2$ and CeRu$_2$. 
We report here 
the results of an extensive investigation of the PE 
phenomenon in a single crystal sample of 
Ca$_3$Rh$_4$Sn$_{13}$ through temperature and magnetic field 
dependence of ac susceptibility and dc magnetization measurements. 
Our results indeed reveal the existence of a characteristic 
 structure, nearly 
identical to that observed earlier in CeRu$_2$ and 2$H$-NbSe$_2$ 
\cite{r23}. The PE region in Ca$_3$Rh$_4$Sn$_{13}$ spans two sharp transitions located at its $onset$ 
and $peak$ positions and a characteristic 
hysteretic effect is observed on thermal cycling across 
the transition located at the onset. Prior to the peak 
position of the PE, the state of the vortex array and its 
current carrying capacity $J_c$ depends on its 
thermomagnetic history, i.e., $J_c$ in the vortex array 
created in the field cooled (FC) manner is much larger 
than that in the zero field cooled (ZFC) manner\cite{r25,r26}. The 
isothermal dc magnetization hysteresis experiments also 
reveal a characteristic path dependence in $J_c$. The new 
results being reported here, therefore, support the generic 
nature of process of disordering of FLL 
under the combined influence of quenched random disorder 
(pinning) and thermal fluctuations.

\section{EXPERIMENTAL DETAILS}
\label{sec:EXPT}

The single crystal of Ca$_3$Rh$_4$Sn$_{13}$
(3.4~x~3.2~x~0.6~mm$^3$, 48.1~mg) used in the present studies 
is from the same batch, grown by the tin flux method and 
utilized by Tomy et al \cite{r24}. The $T_c(0)$ of 
Ca$_3$Rh$_4$Sn$_{13}$ is 8.18~K and it crystallizes in 
the Phase-1 structure (primitive cubic). The ac 
susceptibility measurements in superposed dc magnetic 
fields have been performed using a well-shielded home built 
ac susceptometer \cite{r27}. The ac and dc fields 
are co-axial and the sample is placed in such a way that 
one of its principal axis (cube edge) is always aligned 
parallel with the field (i.e., $H$~//~[001]). The ac 
measurements were usually made at a frequency of 
211~Hz and with an ac amplitude of 1~Oe (r.m.s.). The dc 
magnetization data were obtained on a commercial SQUID 
magnetometer (Quantum Design Inc., U.S.A, Model MPMS 5),  
but using a new procedure 
designated as $Half~Scan~Technique$ 
 (HST) by Ravikumar et al \cite{r28}. The 
magnetization values obtained via HST minimize the 
artifacts arising due to the sample movement through the 
inhomogeneous magnetic field in the SQUID magnetometer 
and are independent of the choice of the scan length. 
Tomy et al \cite{r24} had earlier obtained dc 
magnetization values in their crystal of 
Ca$_3$Rh$_4$Sn$_{13}$ with (full) scan length of 2~cm and 
using the usual full scan (FS) analysis procedure as 
prescribed by the MPMS software. Ravikumar et 
al \cite{r28} have focussed attention on to the 
differences in the magnetization values in the PE region 
of CeRu$_2$ and 2$H$-NbSe$_2$ obtained by HST and FS 
analysis. We have noted similar differences for 
Ca$_3$Rh$_4$Sn$_{13}$ crystal as well and have, therefore, 
used the HST procedure to acquire the dc magnetization data 
reported here.

\section{EXPERIMENTAL RESULTS}
\label{sec:RD} 

\subsection{MANIFESTATION OF PEAK EFFECT VIA ISOFIELD AC 
SUSCEPTIBILITY AND ISOTHERMAL DC MAGNETIZATION 
MEASUREMENTS}

I. ISOFIELD AC SUSCEPTIBILITY MEASUREMENTS

Figures 1(a), 1(b), and 1(c) show the temperature 
dependence of the real part of the ac susceptibility 
($\chi\prime(T)$) for a Ca$_3$Rh$_4$Sn$_{13}$ crystal  in 
various dc bias fields maintained parallel to the [001] 
plane of the crystal. The FLL was 
prepared in the zero field cooled (ZFC) mode (i.e., the 
superconducting sample was initially cooled to the lowest 
temperature in zero magnetic field and subsequently a 
given value of the dc field was applied). The screening 
response was then measured while warming up the sample. 
Figure 1(a) shows the typical variation of 
$\chi\prime(T)$ in low applied 
fields (0~-~2.5 kOe). In zero 
field, $\chi\prime(T)$ response shows perfect screening 
 ($\chi\prime\approx$ -1) at 
low temperatures and a sharp transition towards normal state near the 
superconducting transition temperature $T_c(0)$. As the 
applied field increases, the superconducting to
 normal transition broadens and 
$\chi\prime(T)$ rises monotonically to its normal state 
value at $T_c(H)$. However, for an applied dc field of 
3.5~kOe, as shown in the
 top panel of Fig.~1(b), this monotonic behaviour 
is interrupted by an anomalous dip in  
$\chi\prime(T)$ at a temperature denoted as  $T~=~T_p$ 
(=~7.43~K~at~3.5~kOe). 

Within the Bean's critical state model (CSM) 
\cite{r29,r15}, $\chi\prime(T)$ can be approximated as, 
\begin{equation}
\chi\prime(T)\sim-~1 + (\alpha h_{ac}/J_c),
\end{equation}
where $\alpha$ is a geometry dependent factor, $h_{ac}$ is the 
applied ac field and $J_c$ is the critical current 
density. The above relation implies that the dip in 
$\chi\prime(T)$ is a consequence of enhancement in 
$J_c$, the ubiquitous Peak Effect. As the dc bias field 
is increased further (e.g. to $H_{dc}$~=~5~kOe), the tiny dip in the 
 top panel of Fig.~1(b) transforms
 to a double peak structure, 
as shown in the bottom panel of Fig.~1(b). This two-dip 
structure becomes progressively more prominent as the dc field 
increases and the peak effect phenomenon eventually comprises two very sharp 
($\le$~5~mK width) first order like transitions as 
implied by the data in Fig.~1(c). 
For instance, in a field of 10 kOe in Fig.~1(c),
 $\chi\prime(T)$ displays a sharp ($\le$~5 mK)
 dip at T = $T_{pl}$ (the so called $onset$
 temperature of the peak effect), followed by another sharp
 dip at T = $T_p$ (the $peak$ temperature). 
 Above $T_p$, the $\chi\prime(T)$ response shows a rapid recovery 
towards the zero value.
 The occurrence of sharp changes in $J_c$ at the $onset$ and the $peak$ 
positions of the PE were not apparent in isothermal $\chi\prime (H)$ scans
performed earlier by Tomy et al \cite{r24} in the same crystal. However, the
 $T_p(H)$ values observed in the present isofield
 measurements show good agreement with $H_p(T)$ 
values reported by Tomy et al \cite{r24}.   
It is interesting to note
how rapidly the PE regime 
evolves with increasing field. A more than ten-fold 
increase can be noted in the
 extent of the anomalous dip in 
 $\chi\prime(T)$ values characterising the PE 
between $H_{dc}$~=~3.5~kOe and 
5~kOe. Also, the recovery of $\chi\prime(T)$ 
 above $T_p$ to its normal state value 
becomes steeper as the field 
increases (cf. plots in Figs.1(b) and 1(c)), 
thus indicating the sharpness of the
 transfrmation in the state of vortex matter occurring across 
the peak position of the PE.

   The inset of Fig.~1(c) 
summarizes $T_{pl}$ and $T_p$ values, plotted in the 
$(H,T)$ phase diagram, for various dc bias fields. Below 
5~kOe, one can hardly distinguish between $T_{pl}$ and $T_p$ 
values of the PE, and 
$T_{pl}$ and $T_p$ curves 
 appear to merge at a multicritical point corresponding to 
$T_p(H)$/$T_c(0)$~$\sim$~0.9 in the 
phase diagram. The lowest field down to which
 a residual 
signature of a peak could be ascertained 
 in our data is 3.5~kOe which 
corresponds to a 
$T_p(H)/T_c(0)$ value of about 0.91.

II. ISOTHERMAL DC MAGNETIZATION RESULTS

In order to further explore the different facets of PE observed in the ac 
susceptibility measurements, isothermal dc 
magnetization hysteresis measurements were also performed on the 
same crystal. The
 results at two 
 temperatures are shown in Fig.~2. Magnetization 
curves were recorded for increasing as well as decreasing 
 field cycles over a total scan length of 4~cm using 
the new HST as well as the 
conventional FS method
 on a SQUID magnetometer. The magnetization 
data recorded at 4.5~K shows a clear 
hysteresis in the field interval
 15~-19~kOe between the forward($H\uparrow$) and the 
reverse($H\downarrow$) legs of the field sweep. According to the Bean's 
critical state model (CSM) \cite{r29}, this hysteresis in 
magnetization $\Delta M(H)$~=~$M(H\uparrow) - M(H\downarrow$) 
provides a measure of the macroscopic critical current 
density $J_c(H)$ and is a distinct indicator of the occurrence
 of the PE. 
The field where the magnetization hysteresis bubble is widest 
identifies the peak field $H_p$ and the collapse of 
hysteresis locates the irreversibility field $H_{irr}$, 
above which the critical current density falls
 below the measurable limit of the method used (see 
Fig.~2).

   In the main panel of Fig.~2 at 4.5 K, it is apparent that 
the hysteresis width in the peak effect region measured by the 
HST is significantly larger (and the hysteresis bubble
is much more symmetric) than 
that measured using the conventional FS procedure. It is 
also to be noted that the use of HST results in measurable values of 
magnetization hysteresis below the onset of the PE, whereas the 
 FS method results in the near absence of 
magnetization hysteresis in the same field region. The 
observed distinction is important in the sense that the isofield ac 
susceptibility measurements show diamagnetic $\chi\prime(T)$ 
response (see Figs 1(a) to 1(c)) prior to the arrival of 
PE regime, which clearly demands 
that the dc magnetization should not be reversible prior
to PE (i.e., $J_c~\ne~0$). These results demonstrate 
the efficacy and the necessity of the HST employed in our isothermal dc 
magnetization measurements. To illustrate the point even further, 
the two insets of Fig.~2 show 
a comparison of magnetization hysteresis data obtained by HST and
 FS method at 6.1~K, where the peak field is expected to be about 
 10~kOe (cf. Fig.~1(c)).
A clear signature of the PE can 
be observed in dc magnetization measurements only if the 
$half~scan~technique$ is used.
   
 Even while the PE hysteretic bubble recorded with HST
  at 4.5~K is more symmetric than that recorded with
 FS method, and undoubtedly the proper technique to utilise, nevertheless the difference in fields marking the onset and
  offset of PE in the forward and reverse legs of the bubble
 persists, i.e., $ H_{pl}^f ~>~H_{pl}^r$, i.e., this feature is not an artifact of the method used but is a characteristic property of the system. Such 
 differences in the $onset$ ($ H_{pl}^f$) and the $offset$ ($H_{pl}^r$) fields 
of the PE have been 
widely noticed in several low $T_c$ and 
 high $T_c$ systems, like, UPd$_2$Al$_3$\cite{r3}, 
CeRu$_2$\cite{r2}, 2$H$-NbSe$_2$\cite{r23}, 
Yb$_3$Rh$_4$Sn$_{13}$\cite{r12,r13}, YBa$_2$Cu$_3$O$_7$\cite{r30}, etc.,
  and have led to 
the proposition that the onset of the PE is akin to a 
first order phase transition\cite{r2,r3,r6,r28,r30}. A first
 order change allows for the possibility of the existence
 of thermal hysteresis and the path dependence in the values of
 some physical variables, which in our case is the current density $J_c$.

\subsection{THE DISORDER AND THE HISTORY DEPENDENCE OF THE 
MACROSCOPIC CRITICAL CURRENT DENSITY $J_c$}

I. ISOFIELD AC SUSCEPTIBILITY STUDY:

We have explored the 
thermomagnetic history effects in $J_c$ in the PE
 region through the temperature dependent 
 ac susceptibility measurements. Two different sample 
histories, viz., the ZFC and the FC,
 commonly associated with disordered magnets 
such as spin glasses, were investigated. 
In the FC mode, the sample was rapidly cooled down from the normal 
state in the presence of an applied dc field and then the 
measurements were performed while warming up the FLL 
( i.e., the field-cooled warm up (FCW) mode). Figure~3 displays 
the 
typical response of $\chi\prime(T)$ for 
two thermomagnetic histories in a field of 10~kOe in  
Ca$_3$Rh$_4$Sn$_{13}$. The FLL states in the 
FCW mode can be seen to give larger diamagnetic screening response as compared 
to those produced via 
ZFC mode over a wide range of temperature, up to the peak position of the PE. 

   Since the macroscopic current density 
$J_c$ is directly related to the $\chi\prime(T)$ 
response (Eq.1), the differences in $\chi\prime(T)$ responses
between the ZFC and FCW modes reveal the history 
dependence of the macroscopic critical current density 
$J_c$ in those modes, i.e., $J_c^{FCW}(T)~>~J_c^{ZFC}(T)$
 for $T~<~ T_p$. In an  
 earlier report, Banerjee et al \cite{r23} have 
exhibited the same kind of history dependence in single crystals of CeRu$_2$ 
and 2$H$-NbSe$_2$. Also, from a small angle neutron 
scattering study, Huxley et al \cite{r31} have surmised that the FC state
 in CeRu$_2$ comprises more finely divided
 regions of correlated lattice as compared to those in the ZFC state. The 
difference in ZFC and FCW responses was finally found to 
disappear at the peak position of the PE, somewhat
 akin to the 
magnetic responses in spin glasses where the ZFC and FC 
magnetization curves merge at the 
spin glass transition temperature $T_g$ \cite{r32}. 
Unlike the spin glasses, where the glass temperature $T_g$ could 
  display measurable differences on variation of the frequency and the 
amplitude of the ac field, at least in the 
 vortex system in Ca$_3$Rh$_4$Sn$_{13}$ (see Fig. 4), the onset ($T_{pl}$) and the peak ($T_p$) temperatures of PE do not vary with the 
 change in frequency and/or amplitude of the ac field. Figures 4(a) and
  4(b) display the $\chi\prime(T)$ response in ZFC mode at 10~kOe, when 
frequency is changed from 211~Hz to 21~Hz (ac amplitude is kept constant) and 
the ac amplitude is changed from 1~Oe to 3.5~Oe (r.m.s) (ac frequency is kept 
constant). The observed robust independence of $T_{pl}$ and $T_p$ on the amplitude and especially on the frequency, further attests to the first order nature of the transformations and the consequent absence of pretransitional fluctuation effects, such as, critical slowing down, etc..\\  

II. ISOTHERMAL DC MAGNETIZATION DATA:

To further elucidate the thermomagnetic history effects in $J_c(H,T)$ in  
Ca$_3$Rh$_4$Sn$_{13}$, Fig.~5 summarises the magnetization hysteresis curves 
recorded after obtaining the vortex states in FC mode at fields lying well below 
as well as within the PE region at $T$~=~4.5~K. As per CSM description of 
magnetization hysteresis data  
\cite{r33}, when the critical current density is uniquely prescribed for a 
vortex state at a given $H$ and $T$, the 
single-valuedness of $J_c(H)$ demands that all the magnetization 
values obtained along various paths with different 
thermomagnetic histories should lie within the envelope loop 
defined by the forward and the reverse branches of the 
magnetization curve obtained in isothermal dc magnetization 
measurements \cite{r33,r34}. The dc 
magnetization data illustrated in Fig.5 go beyond
 the above description and 
elucidate the multi-valued nature of $J_c(H)$. 
Figure 5 shows that as the sample 
is cooled in a pre-selected magnetic field ($H_i~<~H_p$) 
to the required temperature (FC mode) and the 
magnetization is recorded as a function of increasing/decreasing fields, the 
initial 
magnetization values (respectively) $overshoot$ the forward/reverse 
magnetization envelope curves. On further 
increasing/decreasing the field, the magnetization values 
fall sharply and merge into the usual envelope loop
(to be identified by the thin continuous lines in Fig. 5).  The 
$overshooting$ by the initial magnetization values clearly indicates that 
the $J_c$ value at a given field $H$ in the FC state is 
higher than that in the ZFC state, in good 
agreement with the isofield ac susceptibility 
measurements discussed earlier. The observation that the FC magnetization 
curves 
eventually merge into the ZFC magnetization curve 
(i.e., the envelope loop)  implies that the more strongly pinned FC 
vortex state heals to a more ordered ZFC state as the 
vortex state adjusts to a large enough change $\Delta H$ in the external dc field\cite{r35}. This change 
$\Delta H$ in the dc field required to anneal a given FC state to a neighbouring 
ZFC like state also 
increases as $H$ approaches the peak field $H_p$, where  
the lattice is nearly amorphous\cite{r7} ($\Delta H$ varies from $\sim$~2~kOe in the PE regime to $\sim$~80~Oe down to very low field far away from the PE regime). The $overshooting$ by the FC 
magnetization curve is however absent for field 
cooling in a field $H_i~>~H_p$ (see Fig.~5). 
In such a  case, the FC magnetization 
curve readily merges with the reverse envelope curve, since the 
history effect in $J_c$ disappears at the peak position of the PE, as seen 
earlier in the temperature dependent ac susceptibility data as well.\\

\subsection{CYCLINGS ACROSS THE ONSET AND THE PEAK
POSITIONS OF PE: EVIDENCE FOR SHATTERING OF THE FLL}

I. THERMAL CYCLINGS DURING ISOFIELD AC SUSCEPTIBILITY MEASUREMENTS: \\ 

	A deeper insight into the path dependence in 
$\chi\prime(T)$ can be brought out by  performing 
thermal cyclings  across the $onset$ and the $peak$ positions
  of PE as it yields striking evidence of the stepwise 
pulverization of the FLL \cite{r23}.\\

  The FLL prepared in the ZFC mode, 
after applying the required field was
 warmed up to three pre-selected 
temperatures,  
$T_I$, $T_{II}$ and $T_{III}$, such that,
 (i) 
$T_I~<~T_{pl}$, (ii) $T_{pl}~<~T_{II}~<~T_p$ and (iii)
 $T_{p}~<~T_{III}~<~T_c$. The $\chi\prime(T)$ responses 
were then recorded while cooling down in field (FCC) from the 
above three pre-selected temperatures. The results are 
illustrated in Fig.~6, where the short-dashed and the solid 
lines (with data points omitted) are the responses 
produced in the ZFC and FCW modes, respectively; these were shown earlier in Fig.~3. The cool down 
curves 
from the three pre-selected temperatures 
are represented 
by specific symbols.
The following features are noteworthy in Fig.~6:\\

  (1) For cooling down the sample from a temperature 
less than the onset of peak effect 
($T_I~<~T_{pl}$), the cool down curve retraces the 
ZFC warm-up curve, i.e., the changes
 in the FLL that occur along this path
 are  reversible. \\

 (2) For cooling down from a temperature $T_{II}$, such 
that $T_{pl}~<~T_{II}~<~T_p$, the $\chi\prime(T)$ 
response initially tries to move towards that of the more ordered 
ZFC state, thereby, effectively retracing the 
ZFC response. But, 
suddenly very close to $T~=~T_{pl}$, i.e., 
just above the onset of the PE, it 
undergoes a huge, sharp jump and $\chi\prime(T)$ drops
 precipitously  to a 
value more diamagnetic, and hence, the FLL far 
more disordered than 
the corresponding state during the 
ZFC warm-up cycle. (In fact,
 the $\chi\prime(T)$ response just below $T_{pl}$ (during FCC from $T_{II}$) 
is even more diamagnetic 
and thus 
 the FLL more disordered than the corresponding
 FLL state during the FC warm up cycle.) This sharp jump towards a 
new, highly disordered state reveals a characteristic 
hysteretic behavior across the $T_{pl}$ transition. A
novel feature of this hysteretic response across $T_{pl}$ 
is that it is like an open hysteresis loop, i.e.,  the 
$\chi\prime(T)$ response can
 never by itself (at $T~<~T_{pl}$) recover to the ZFC like 
response. On cooling down further, 
the new $\chi\prime(T)$ curve (FCC from $T_{II}$) 
cuts across the FCW response, thereby, becoming relatively 
less diamagnetic 
than the FCW curve down to the 
lowest temperature ($T<<T_{pl}$).\\

 (3) When the FLL is cooled down from a temperature above the 
peak position of the PE ($T_{III}~>~T_p$), the cool down 
ac susceptibility response initially appears to retrace the FCW 
curve down to $T_p$. But below $T_p$, the $\chi\prime(T)$ 
response becomes more diamagnetic than both the ZFC and FCW 
responses, thereby creating a vortex state with even still larger  
$J_c$ value. Between 
 $T_p$ and $T_{pl}$, 
 this highly disordered state stages a partial 
 recovery. However, near the onset of PE ($\geq$~$T_{pl}$), 
this partially recovered
 disordered state has $J_c$ closer to (but larger than) that of the state obtained 
by cooling down (to T~$\le$~$T_{pl}$) from $T_{II}$; but below $T_{pl}$,
 it undergoes further disordering 
(see the curve FCC from $T_{III}$). It may be noted that enventually
 at $T~<<~T_{pl}$, $\chi\prime(T)$ response asymptotically 
merges with the FCW curve. \\
 
  The above results bear close resemblance with
 the data in 2$H$-NbSe$_2$ and 
 CeRu$_2$ \cite{r23} and reaffirm the
 notion of the stepwise 
shattering process of the FLL across the temperatures $T_{pl}$ and $T_p$.\\

 II. CYCLINGS VIA MINOR HYSTERESIS LOOPS IN AN ISOTHERMAL
 DC MAGNETIZATION EXPERIMENT :\\ 

Another interesting manifestation of the path dependent behavior across the onset position of PE can be 
seen in the isothermal dc magnetization hysteresis data via
 a characteristic feature in the so called 
  minor hysteresis loops, as reported first by
 Roy and Chaddah \cite{r2} in several samples of CeRu$_2$
  and its derivatives\cite{r6,r36}. It was pointed out \cite{r2,r6,r30,r36} that 
 the minor loops initiated from fields 
 lying in between the onset and the peak fields
 of PE (i.e., for $H_{pl}~<~H~<~H_p$), saturate
 without merging with the reverse envelope loop. We show
 in Figs.~7(a) and 7(b) the behaviour
 of minor magnetization curves initiated from fields lying
 in between $H_{pl}^f$ and $H_p$ at 4.5~K 
 in Ca$_3$Rh$_4$Sn$_{13}$. The expanded data in the 
 main panel of Fig.~7(b) clearly elucidates that the \rm{saturated} 
 (i.e., the highest) values of the minor loops do not meet
 the reverse envelope curve (identified by the 
 continuous line). The inset panel of Fig.~7(b) shows how the 
 minor loops eventually merge into the reverse envelope near 
 $H_{pl}^r$. Note that the difference between the saturated 
 value of a minor curve and the magnetization
 value on the reverse envelope curve decreases as the 
field from which the minor loop is initiated
 approaches the peak field $H_p$. For fields above $H_p$, the minor curves
 readily merge into the reverse envelope curve.\\

   Notionally, 
 above $H_p$, the $J_c(H)$ values on the forward and reverse legs of
 the envelope hysteresis loop  are just phase reversed, i.e., their
 magnitudes do not display path dependence. An
 estimate of the current density values along the forward/reverse
 portion of the envelope loop between  $H_{pl}^f$/$H_{pl}^r$
  and $H_p$ can be made by examining the respective half
 widths of the magnetization hysteresis loop\cite{r35}. The magnetization
 data of Fig.~7(b) leads to the inference that $J_c^f(H)~<~J_c^r(H)$
 for $H_{pl}~<~H~<~H_p$. The 
 inequality $H_{pl}^f~<~H_{pl}^r$, and the distinct 
identities of the different members of the family of minor loops
 in the field interval $H_{pl}^r$ and $H_{pl}^f$ vividly exemplify the 
irreversibility and the path dependence in the physical phenomenon
 that occurs across the onset position of the PE along the forward leg.\\

\section{DISCUSSION}
\label{sec:DIS}

\subsection{SUPERCONDUCTING PARAMETERS OF Ca$_3$Rh$_4$Sn$_{13}$ :}

Ca$_3$Rh$_4$Sn$_{13}$ is an isotropic superconductor, 
well described under the framework of phenomenological Ginzburg-Landau (GL) theory in the isotropic limit. Table 1 summarizes the values of various 
physical parameters of Ca$_3$Rh$_4$Sn$_{13}$ at a temperature of 4.5~K  obtained\cite{r37} from the ac susceptibility and dc magnetization data using  the GL theory. The dimensionless Ginzburg-Landau 
parameter $\kappa$, defined as the ratio of the two 
characteristic lengths $\lambda$ and $\xi$, obtained for 
this isotropic  low $T_c$ superconductor is $\approx$~17.5,  
whereas the coherence length $\xi$ and the penetration 
depth $\lambda$ at 4.5 K are $\sim$~130~$\AA$ and 
$\sim$~2270~$\AA$, respectively.  The presence of 
disorder, in the forms of thermal fluctuations (dynamic) and the quenched 
random pinning centres (static), governs the statics and the 
dynamics of the vortex array. The importance of these 
two sources of disorder can be quantified\cite{r38} by the
values of Ginzburg 
number ($G_i$), defined as  
$G_i~=~(1/2)(k_BT_c/H_c^2\xi^3\epsilon)^2$, and the ratio $j_c$/$j_0$,
 where $j_c$ 
and $j_0$ denote the depinning and depairing current densities, 
respectively. In order to conveniently observe the melting
 or the amorphization transition of the vortex lattice in any 
type II superconductor, the following two criteria
 must be fulfilled. The 
system must have an appreciably large value of Ginzburg number 
$G_i$, which guarantees an adequate separation of the vortex 
melting/amorphization 
curve from the $H_{c2}$ line. The system must 
have sufficiently wide very weakly pinned region well  
below the superconductor-normal phase boundary, which could
 help to identify the occurrence of the PE in the 
magnetization experiments, which is a 
reliable signature of the process of amorphization of
 vortex lattice.  The Ginzburg number $G_i$, which 
 measures the relative size of the minimal ($T$~=~0) 
condensation energy $H_c$$^2$(0)$\epsilon$$\xi$$^3$(0) 
within a coherence volume and the critical temperature $T_c$, 
for Ca$_3$Rh$_4$Sn$_{13}$ is $\sim$~10$^{-7}$. This value 
is not too large\cite{r1,r38} indicating that Ca$_3$Rh$_4$Sn$_{13}$ system  
has a rather narrow critical 
fluctuation region in $(H,T)$ phase space, where the vortex 
lattice can undergo melting transition 
due to thermal fluctuations alone. 
However, the ratio $j_c$/$j_0$ 
is $\sim$~7$\times$~10$^{-5}$ for weakly 
pinned Ca$_3$Rh$_4$Sn$_{13}$ crystals, which 
is few orders of magnitude lower than that in the 
typical crystals of high $T_c$ cuprate systems\cite{r1,r38}. 
Such weak pinning in Ca$_3$Rh$_4$Sn$_{13}$ crystals
can be conjectured to be a consequence of the absence of 
generic extended pinning sites, thereby leading to the 
assumption that pinning is being mainly
 structural and provided by the point defects in this system. 
The weak pinning situation is very helpful as the structural and
the dynamical behaviour of the vortex array can be 
studied within the framework of the 
 Larkin-Ovchinnikov (LO) collective pinning 
theory\cite{r39,r40}. One can therefore safely assume 
that FLL in Ca$_3$Rh$_4$Sn$_{13}$ is well-formed with  
large Larkin domains.  The small $j_c$/$j_0$ 
ratio makes Ca$_3$Rh$_4$Sn$_{13}$ an attractive candidate for the 
study of pristine phase transitions\cite{r38,r41,r42,r43} 
 of vortex matter in disorder-free context, like,
the FLL amorphization transition through the 
observation of the phenomenon of peak effect\cite{r1}.\\

\subsection{COLLECTIVE PINNING DESCRIPTION}

The motivation to invoke LO description\cite{r39}
 for explaining the observed behaviour in Ca$_3$Rh$_4$Sn$_{13}$ is
 supported by the field dependence of critical current density
 in it. $J_c(H)$ can be estimated within the prescription
 of critical state model\cite{r29} using magnetization hysteresis
 data of Fig.~2 and the 
knowledge of the dimensions of the crystal\cite{r33}. Figure 8 shows $J_c(H)$
 vs $H$ behaviour at 4.5 K on a log-log plot. In the ZFC mode, $J_c^{ZFC}$ vs $H$ displays a linear variation in Fig.~8, which amounts to a power law dependence for $J_c(H)$ prior to reaching the PE region at $H_{pl}$. Such a power law
 dependence is 
often taken to be a signature of collectively pinned elastic
 solid\cite{r44}. Figure~8 also includes the current density in the FC state 
estimated
 using the saturated values of minor hysteresis curves initiated from different
 $M_{FC}(H)$ values. 
The difference between the highest magnetization value on the
  minor hysteresis curve initiated from 
a given $H$ and the notional equilibrium magnetization
  value could be taken as a measure of $J_c^{FC}(H)$\cite{r35}.
It is apparent from Fig.8 that the difference between $J_c^{FC}(H)$
 and $J_c^{ZFC}(H)$ values decreases as $H$ decreases such that at a field
 of about 4~kOe, one cannot distinguish between the FC and ZFC vortex states
 on the basis of their critical current density values.
 A similar trend was noted\cite{r26} in a crystal of 2$H$-NbSe$_2$, where
the difference between the $J_c$ values in 
FC and the ZFC states ceased at about 1 kOe. 
It is
noteworthy to recall here that in 
another study\cite{r45} 
on a variety of 2$H$-NbSe$_2$ 
  crystals, it was demonstrated that the structure 
in the PE and thermomagnetic history effects 
in $\chi\prime(T)$
 data correlate with effective pinning in a manner
 that the first drop in
 $\chi\prime(T)$ at $T_{pl}$ is pinning 
induced. For a given amount 
 of quenched disorder, the effective pinning 
increases as $H$ increases.
 We explored this notion in Ca$_3$Rh$_4$Sn$_{13}$ as well via the study of 
 $\chi\prime(T)$ response in ZFC and FC states in a field of 5.3~kOe and its
 comparison with the data recorded in higher
 dc fields. Figure 9 
summarizes the $\chi\prime(T)$ data recorded 
at 5.3 kOe. It 
can be noted that there is only a tiny 
difference between ZFC and FC 
curves at $T~<<~T_{pl}$. The inset in Fig.~9 shows
 how the difference
 between the ZFC and the FC data, along with the 
structure in the PE
 across $T_{pl}$ and $T_p$, has become
 less prominent at 5.3 kOe as compared to 
 the behaviour displayed in Fig.~1(c) for 10 kOe.\\
 
  Within the frame work of 
LO model, the presence of quenched random 
disorder in the elastic FLL  destroys the perfect long range 
order of the FLL through a competition between the 
interaction (i.e., the rigidity of the FLL) and 
the randomness leading to a limited spatial region within which
 the FLL remains well formed. The 
macroscopic critical current density $J_c$ in the 
presence of pinning is given by the pinning force 
equation\cite{r39,r40}: 
\begin{equation}
F_p=J_cH=(W/V_c)^{1/2}=[n_p~<f_p^2>/R_c^2L_c]^{1/2},
\end{equation}
where $W$ and $V_c$ represent the pinning parameter
and the correlation volume of a Larkin domain within which the FLL is 
undistorted 
and well correlated despite the presence of pinning. In the above equation, 
$n_p$ is the density of pinning centers (pins), $f_p$ 
is the elementary pinning interaction proportional to the 
condensation energy and $R_c$ and $L_c$ 
are collective pinning radial and longitudinal 
correlation lengths. The 
correlation lengths depend on both the elasticity of the 
FLL as well as pinning, characterized by the parameter W, and are 
given by,

\begin{eqnarray}
R_c&=&[c_{44}^{1/2}c_{66}^{3/2}r_f]/W \nonumber\\
~~{\rm and}~~L_c&=&[c_{44}/c_{66}]^{1/2}R_c,
\end{eqnarray}
where $c_{44}$ and $c_{66}$ are the tilt and 
the shear moduli of the FLL, respectively, and $r_f$ is 
the range of interaction of the pinning force. In 
the LO theory, the pinning interaction can only decrease 
with increasing $T$, which corresponds only to a monotonic 
decrease in $J_c$ with increasing $T$. Thus, the anomalous 
increase in $J_c$ at the PE can be 
rationalized in the LO scenario 
only by an introduction of an anomalous 
sudden drop in the Larkin volume $V_c$. The 
two discontinuous jumps in the current
 density $J_c$ at $T_{pl}$ and $T_p$ in 
$\chi\prime(T)$ data, therefore, imply a sudden shrinkage of 
$V_c$ at $T_{pl}$ and $T_p$. This picture proposes that 
for a  sample with substantial pinning, the competition 
between interaction and disorder leads to some threshold 
behaviours. When the pinning energy overcomes the elastic 
energy of the FLL, which is continuously softened 
due to increase in $T$, a disorder induced transition  
transforms the FLL from an elastic medium to a 
plastically deformed vortex state with a proliferation of 
topological defects (dislocations) at the onset of PE 
corresponding to the first jump in the ac susceptibility 
response (at $T_{pl}$)\cite{r23}. At a higher temperature, the 
thermal energy overcomes the elastic energy and the 
pinning induced shattering transition\cite{r23,r45}
 produces a complete 
amorphization of the vortex solid at the peak position of 
the PE, which could be identified with the second jump in the 
diamagnetic response at $T_p$.\\

  The crucial support for the stepwise fracturing 
of the FLL comes 
from the thermal cycling experiments performed on a
well ordered FLL generated in a ZFC manner (see Fig.~6). 
When the temperature increases towards the 
PE region, the FLL softens, the energy needed to create 
the dislocations decreases and the lattice spontaneously 
fractures at $T_{pl}$. While cooling down from 
a temperature within the 
PE pocket, i.e., between $T_{pl}$ and $T_p$, the FLL is 
cooled from a partially fractured, plastically deformed 
state. So the FLL initially tries to heal towards the 
well ordered ZFC state thereby retracing the ZFC 
response. But, as the temperature decreases progressively
 towards $T_{pl}$, the decrease 
in strength of thermal fluctuations start to stiffen the FLL thereby 
building up stresses. The system thus fails to drive out the 
dislocations in order to heal back to the ZFC state. 
Instead, it shatters further to release the stresses and 
reaches a new metastable state. This characteristic feature also 
yields an open hysteresis loop, i.e., after the 
$\chi\prime(T)$ response has reached the new metastable 
state, it is unable to recover to the ZFC like 
response by itself. To reach back to the ZFC state from this 
metastable state, one needs to shake the FLL vigorously 
by an external driving force\cite{r35}. This characteristic 
hysteretic behaviour across the $T_{pl}$ transition is 
definitely novel; it is not usually seen across a typical first order 
(melting) transition in the absence of disorder.

The differences between the vortex states formed during
  cool down from $T>T_p$ and those evolving during FCW mode
 probably originate from the slow dynamics\cite{r1,r23,r26} of 
the order- disorder transformation
 across $T_{pl}$ and $T_p$.
We believe that above $T_p$, the FLL 
transforms into a completely pulverized state,  
which is so disordered that the FLL 
correlations beyond first few 
 nearest neighbour inter-vortex spacings are immaterial. Thus,  
if the FLL is cooled down from such a disordered
 state (i.e., $T~>~T_p$), one cools in 
liquid-like correlations and the FLL remains in an 
amorphous state (supercooled state) down to a much lower 
temperature
 (than that during field cooled warm up cycle). 
The above scenario
 leads us to infer that at 
$T_{pl}$, the FLL transforms from a nearly defect-free 
ordered lattice, such as, a Bragg glass phase 
\cite{r41,r42} to a highly defective plastically deformed 
lattice \cite{r41,r42,r43,r46,r47,r48,r49} with full of 
topological defects analogous to a vortex glass phase. 
Then at $T_p$, 
the vortex glass phase further transforms into a 
completely amorphous but pinned phase 
($J_c~\neq$~0), above which the lattice is no longer 
correlated and loses its history and memory. The vanishing of the
 pinning occurs at an even higher temperature 
marked as irreversibility temperature $T_{irr}$.

The path dependence in $\chi\prime(T)$ response for T~$\ll~T_p$ 
is a thought provoking result. It had been shown 
by M$\ddot{\rm {u}}$ller, Takashige and Bednorz\cite{r50} 
that the $M_{ZFC}$ differed from 
$M_{FC}$ up to $T_{irr}$, and hence one could produce a 
hierarchy of states whose bulk magnetization values lie 
in between $M_{ZFC}$ and $M_{FC}$. In the ZFC state, one typically  
establishes the critical state with critical current density 
$J_c(H)$. The states with magnetization lying in 
between $M_{ZFC}$ and $M_{ZFC}$ are in sub-critical 
state with current densities $J(H)~\le~J_c(H)$\cite{r51}. 
At $T_{irr}$, 
$J_c(H)$ is considered to approach the depinning limit ($J_c$$\to$0). 
The magnetic shielding response as measured via $\chi\prime(H,T)$ 
of all the states whose 
magnetization values lie in between $M_{ZFC}$ and $M_{FC}$ is 
dictated by $J_c(H)$ and is nearly identical, i.e., 
$\chi\prime(T)$ does not display any path dependence 
though the magnetization values are path dependent. On 
the contrary, we witness a thermomagnetic history dependence 
in $\chi\prime(T)$ responses in Ca$_3$Rh$_4$Sn$_{13}$ 
below the peak position of the PE. The FLL states prepared via 
different paths in the ($H$,$T$) phase space produce 
different $\chi\prime(T)$ responses. The quenched 
random disorder appears to pin the vortices into different 
configurations  
while preparing the FLL via different routes in the ($H$,$T$) 
plane and this results in the history dependence in the 
screening responses of the FLL. This path dependence in 
$\chi\prime(T)$ further reflects the path dependence in 
$J_c$ \cite{r25,r35}, which in turn gives the path dependence 
of the correlation (Larkin) volume of the FLL. This leads  
us to infer that the ZFC state is perhaps the most ordered state with 
the lowest $J_c$ and the slowly cooled FC state as the most disordered 
state with the largest $J_c$. Hence, one can produce a  
hierarchy of metastable states 
whose $J_c$ values lie in between $J_c^{ZFC}$ and 
$J_c^{FC}$. The history dependence of $\chi\prime(T)$ 
responses disappears at the peak position ($T_p$) of the 
PE, presumably due to the fact that the FLL prepared 
in any manner almost completely amorphizes at the 
peak position and the vortex system $loses~order$ in 
equilibrium.\\ 

\subsection{ DETERMINATION  OF  $R_c$  AND $L_c$ 
BY COLLECTIVE PINNING ANALYSIS}

Within the LO framework of collective pinning, 
the dimensionality (D) of the collective pinning is determined by the relative 
magnitude of longitudinal correlation length $L_c$ of a Larkin domain 
with respect to the thickness $d$ of the sample\cite{r38,r39,r40}. Thus a  
transition from 3D to 2D behaviour is predicted only when $L_c$ $\ge$ $d/2$. 
The estimated value of $L_c(0)$ from 
the $j_0$/$j_c$ ratio for Ca$_3$Rh$_4$Sn$_{13}$  
is much smaller than $d/2$ which indicates that the 
collective pinning in this system is 3D in nature. We estimate the longitudinal 
and transverse correlation lengths $L_c(H)$ and $R_c(H)$ respectively of the 
Larkin volume, at 4.5 K, using various 
superconducting parameters calculated from isotropic 
GL theory
(cf. Table 1), within the framework of 3D collective pinning. 

The pinning parameter $W(H,T)$ which takes into account both the density 
as well as the strength of pins, can be written in a separable form
 as\cite{r40}~: 
\begin{equation}
W(H,T)=W_0(T)f(b),
\end{equation}
where the
field dependent part $f(b)$ of the pinning strength $W$ is determined by the relation
\begin{equation}
f(b)=b(1-b)^2 ,
\end{equation}
where $b$ is the reduced field given by $b = B/\mu_0H_{c2}$. $W_0$ can be determined from the expression for single vortex pinning\cite{r38}, i.e.,
\begin{equation}
W_0=L_c(0)j_c^2(0)\phi_0\mu_0 H_{c2}(T),
\end{equation}
where $\phi(0)$ is the flux quantum and $L_c(0)$ is the longitudinal collective pinning length given by, 
\begin{equation}
L_c(0)=\xi(0)[j_{0}(T)/j_c(0)]^{1/2}.
\end{equation}
The ($j_0$(T)/$j_c(0)$) ratio for this
 superconductor is $\sim$$10^{4}$ and the typical values obtained for $L_c$(0) and $W_0$ at 4.5 K are
 1.6~$\mu$m and 6~$\times10^{-7}$ $N^2/m^3$.
Incorporating the expression of $W$ in the LO expression for collective pinning (Eq.~2), one gets,
\begin{equation}
J_cH=(W_0b(1-b)^2/R_c^2L_c)^{1/2}.
\end{equation}
The longitudinal correlation length $L_c$ and the transverse correlation length $R_c$ are again connected by the equation\cite{r40} :
\begin{equation}
L_c=(c_{44}/c_{66})^{1/2}R_c~\approx(2\sqrt{2}/\pi\xi)(b/1-b)^{1/2}R_c^2,
\end{equation}
where $c_{66}$ and $c_{44}$ are the shear and the tilt moduli, respectively, given by, 
\begin{equation}
c_{66}\approx[\phi_0B/16\pi\mu_0\lambda^2](1-b)^2=[\phi_0^2/32\pi^2\mu_0\lambda^2\xi^2]b(1-b)^2 .
\end{equation}
and
\begin{equation}
c_{44}(k_{\perp}, k_z)\approx c_{44}^0/
(1 + \lambda_h^2k_z^2 + \lambda_h^2k_{\perp}^2)\approx c_{44}^0(1 + \lambda_h^2k\perp^2)
\approx c_{44}^0R_c^2/\pi^2\lambda_h^2
\end{equation}
by considering $L_c$~$\gg$~$R_c$ where $k_{\perp}$ and $k_z$ are the wave vectors describing the 
deformation fields normal and parallel to the field direction respectively,  $\lambda_h~=~\lambda/(1-b)^{1/2}$ is the renormalized length scale\cite{r52} of the order parameter $\lambda$ at large induction B and $c_{44}^0~=~B^2/\mu_0$.
Plugging Eq.~9 into Eq.~8, one can obtain the expressions for the two correlatiobn lengths $R_c$ and $L_c$ as,
\begin{equation}
R_c~=~[(W_0b(1-b)^2)/(2\sqrt{2}/\pi\xi(b/(1-b))^{1/2})]^{1/4}[1/J_cH]^{1/2}
\end{equation}
and
\begin{equation}
L_c~=[~2\sqrt{2}/\pi\xi(b/(1-b))^{1/2}]^{1/2}[W_0b(1-b)^{1/2}]^{1/2}[1/J_cH].
\end{equation}

 Figure 10 summarizes the computed data for the field dependence of the ratios 
$R_c/a_0$ and $L_c/d$ for the FLL prepared at 4.5 K in ZFC mode as well as in FC 
mode in the Ca$_3$Rh$_4$Sn$_{13}$ crystal under study. 
For the ZFC mode, both $R_c/a_0$ and $L_c/d$ initially increase with field upto 
the onset field $H_{pl}$ of PE. But, between $H_{pl}$ and $H_p$, they decline
rapidly indicating that the elastic energy decays much faster than the pinning 
energy as $R_c$ and $L_c$ decay faster than the pinning parameter $W$. Near 
the peak field $H_p$ of the PE, $R_c/a_0\to$1 which supports the view that the 
FLL gets completely amorphized as the shear modulus $c_{66}$ goes to zero. 
However, for the FC state, both $R_c/a_0$ and $L_c/d$ show weaker variation with 
$H$ upto 0.8~$H_{pl}$, after which they decrease gradually, reaching the amorphous limit near $H_p$. \\

\subsection{VORTEX PHASE DIAGRAM}

Collating all the ac susceptibility and dc magnetization 
 data (including those from Ref. 24), 
we can finally construct a vortex phase diagram (see Fig.~11) of 
the conventional type~II superconductor 
Ca$_3$Rh$_4$Sn$_{13}$. In Fig. 11, the lower critical field 
line $H_{c1}(T)$, below which the 
superconductor is in the Meissner state, has been drawn from 
an analysis of isothermal dc 
magnetization measurements
 performed at low fields\cite{r37}. The $H_{c2}$ line, 
above which the system is in normal state, has been drawn from 
the values obtained from the ac susceptibility as well as 
dc magnetization data. The regime in the magnetic phase 
diagram enclosed by these two lines gives the Abrikosov 
state or the mixed state of a type II superconductor, where the 
magnetic field penetrates in 
the form of discrete flux quanta which 
interact repulsively among themselves to form the FLL. 
After the discovery of high $T_c$ superconductors, 
M$\ddot{\rm {u}}$eller, Takashige and Bednorz \cite{r50} 
introduced an extra line, designated as an irreversibility line
 in the mixed state of type II superconductors,
 which was thought to mark the predominance of the influence of  
thermal fluctuations on the vortices. In weakly pinned systems, the 
competition and interplay between the static (quenched 
random pinning) and dynamic (thermal fluctuations) 
disorder and the elasticity of the lattice 
creates a richer phase diagram comprising a few more lines 
within the Abrikosov phase. These lines reflect how the nearly 
defect-free (very small $j_c$/$j_0$ ratio) elastically 
deformed vortex lattice undergoes loss of order in steps. 
The onset of the PE, which is 
identified with $T_{pl}(H)$ in ac susceptibility data 
and $H_{pl}(T)$ 
in dc magnetization data is drawn as ($H_{pl}$,$T_{pl}$) line in 
Fig. 11. This line marks the onset of shattering of the FLL as the 
pinning energy overcomes the elastic energy 
and hence, a transition ensues from an elastic solid similar to 
Bragg glass phase to a plastically deformed vortex state 
full of topological defects analogous to a vortex glass 
phase. The shattering of the FLL gets completed at the 
peak position of the PE (marked as ($H_p$,$T_p$)) producing a 
near collapse of the Larkin domain and complete 
amorphization of the FLL, thereby, representing a 
transformation to a highly viscous pinned amorphous state as the elastic energy is completely overcome by the thermal energy. 
If we choose to identify ($H_p$,$T_p$) line in Fig.~11 with the 
FLL melting curve ($B_m$, $T$) relation\cite{r38}:

\begin{equation}
B_m=\beta_m(c_L^4/G_i)H_{c2}(0)(T_c/T)^2[1 - (T/T_c)-
(B_m/H_{c2}(0)]^2,
\end{equation}

\noindent where $\beta_m$, $G_i$ and $H_{c2}$(0) are 
taken as 5.6\cite{r38}, $\sim1.3\times10^{-7}$ and $\sim$4.5~T, respectively
, the Lindemann number $c_L$ is found to be 
$\sim$~0.1. At a higher field $H_{irr}$, the critical 
current density vanishes
as identified by the vanishing of hysteresis in the 
dc magnetization data. The irreversibility line
 marks a crossover (presumably 
dynamic) from a pinned amorphous to an unpinned 
amorphous state as the thermal energy overcomes the pinning 
energy. \\

SUMMARY

In summary, the peak effect, an anomalous upturn in the 
critical current density of superconductors, has been 
investigated in detail for the superconducting system 
Ca$_3$Rh$_4$Sn$_{13}$ via isofield ac and isothermal dc 
magnetization measurements. A structure in the peak effect 
 region, comprising two 
first-order like jumps at the onset ($T_{pl}$) and the 
peak ($T_p$) positions of the PE, has been observed in the  
isofield ac susceptibility measurements. Our ansatz about 
this two-peak structure is that the first 
peak reflects the commencement of 
 a pinning induced 
stepwise shattering of the FLL through the sudden 
shrinkage of the correlation (Larkin) volume $V_c$ at $T_{pl}$\cite{r23,r45}. 
Thermal cycling across the onset temperature $T_{pl}$ 
reveals an open hysteresis loop across it, which is a 
spectacular manifestation of the shattering phenomenon 
of the FLL. The macroscopic current density $J_c$ of 
Ca$_3$Rh$_4$Sn$_{13}$ shows pronounced thermomagnetic 
history dependence below $T_p$, which reveals the role 
played by quenched random disorder on the FLL. The 
 disappearance of history dependence above $T_p$, 
reflects the absence of memory of any previous history  
 and the complete amorphization of the FLL. 
The two-peak structure becomes inconspicious below a certain dc 
field indicating a possible cross over from an interaction 
dominated regime to a pinning dominated regime when lattice constant 
$a_0$ becomes comparable to the penetration depth, which 
measures the range of the electromagnetic interaction between the vortices. Finally, the vortex phase diagram 
of Ca$_3$Rh$_4$Sn$_{13}$ is constructed which 
shows close resemblance to phase diagrams 
drawn earlier for CeRu$_2$ and 2$H$-NbSe$_2$\cite{r23}. This could further 
lead to 
the establishment of a generic phase diagram for all conventional 
low-$T_c$ type~II superconductors in the presence of 
quenched random disorder and thermal fluctuations.\\

\clearpage

\begin{figure} 
\caption{(a),(b),(c) Isofield ac susceptibility data for 
various fixed dc bias fields, 
superimposed 
with an ac field of amplitude 1 Oe (r.m.s) at a frequency of 211 Hz,
maintained parallel to [001] 
plane of a single crystal of Ca$_3$Rh$_4$Sn$_{13}$. 
 Figure 1(a) demonstrates 
that for dc fields of 0 Oe (i.e., earth's field) and 2.5 kOe, $\chi\prime(T)$
 response exhibits monotonic decrease with T. 
However, in the upper panel of Fig.1(b), for a dc field of 3.5 kOe, the
 monotonic decrease of $\chi\prime(T)$ response is 
 interrupted
suddenly, giving rise to 
an anomalous dip just before $T_c(H)$.
 The lower panel of Fig.1(b) 
 sketches how the single dip-like feature transforms to a two peak structure,
occurring at the $onset$ ($T_{pl}$) and the $peak$ ($T_p$) positions of the peak effect,
 as the dc bias field is increased further to 5 kOe. Figure 1(c)
 shows how the structure in the PE regime gets more prominent as one 
progresses to 
 higher dc fields of 7 kOe and 10 kOe. 
The inset of Fig. 1(c) shows the $T_{pl}$ and
 $T_p$ lines, plotted in the thermomagnetic $(H,T)$ phase space.}   
\label{Fig. 1 (a),(b),(c)}
\end{figure}
\begin{figure}

\caption{The main panel shows the isothermal dc magnetization 
hysteresis data at 4.5 K obtained via Quantum Design Inc. SQUID
Magnetometer system over a scan length of 4 cm with the usual full 
Scan (FS) method and a new {\it half scan technique} (HST). 
$H_{pl}^f$ 
and $H_{pl}^r$ identify the field values at which PE notionally commences 
on the forward leg and terminates on the reverse leg, respectively. 
The peak field of the peak effect, the irreversibility field (where
magnetization hysteresis bubble collapses) and the upper critical field are  marked as $H_p$, 
$H_{irr}$ and $H_{c2}$ , respectively. 
The two insets in Fig.2 show the magnetization measurements 
at 6.1 K using HST as well as conventional FS technique. No 
hysteresis bubble in the magnetization data a la PE is observed for the FS method, whereas the HST, which circumvents the problem of field 
inhomogeneity, produces measurable magnetization hysteresis in the  PE regime. Also, 
the main panel of Fig. 2 focusses onto a comparison between the two 
magnetization bubbles measured at 4.5 K by conventional FS method and HST. 
Besides other differences in the two loops, note that the loop via HST 
is open before the onset field $ H_{pl}$.}
\label{Fig. 2}
\end{figure}

\begin{figure}
\caption{Temperature dependence of the in phase ac susceptibility
 data recorded at $H$~=~10 kOe for two thermomagnetic histories, namely,
zero field cooled(ZFC) and field cooled(FC). In the latter case, 
the data were recorded during warm-up cycle (FCW). Note that the
 jumps in $\chi\prime(T)$ are observed at $T_{pl}$ and $T_p$ in both the plots, 
and just above $T_p$, the two curves merge into each other.}
\label{Fig. 3}
\end{figure}

\begin{figure}
\caption{(a),(b) Frequency and amplitude 
dependence of the isofield ac susceptibility
 measurements for a FLL prepared by a dc bias field 
of 10 kOe. Figure 4(a) shows the $\chi\prime(T)$
 responses for two different frequencies of 21 Hz and 211 Hz, 
whereas Fig. 4(b) shows
 a comparison of data obtained at two 
different amplitudes of ac field keeping the frequency invariant.
 Note that $T_{pl}(H)$ and $T_p(H)$ values in H~=~10 kOe do not show measurable 
differences over the limited ranges of the frequency and the amplitude investigated.}
\label{Fig. 4 (a),(b)}
\end{figure}

\begin{figure}
\caption{Magnetization hysteresis curves obtained 
at 4.5 K after field cooling the crystal to
different preselected fields. $M_{FC}(H)$ 
denotes the notional field cooled 
magnetization value at a given $H$. Note that the minor magnetization curves can be obtained by either increasing or decreasing the field from a given $M_{FC}$(H). The field cooled minor magnetization curves so obtained initially 
cut across the envelope magnetization loop (marked by thin continuous line). 
The FC minor curves eventually proceed towards the envelope loop. Note that the
 characteristic behavior which amounts to a FC minor curve of cutting across the envelope curve ceases as $H$
 approaches the peak field $H_p$. Above $H_p$, the FC minor curve readily merges 
into the envelope loop.}
\label{Fig. 5}
\end{figure}

\begin{figure}
\caption{Thermal cycling across the onset ($T_{pl}$)
and the peak ($T_p$) positions of the 
peak effect in
 tempearture dependent ac susceptibility data.
 The short-dashed
 and the solid lines represent the $\chi\prime(T)$ responses
 for the ZFC and the FCW states, respectively, which were
 displayed earlier in Fig.3. $T_I$, $T_{II}$ and $T_{III}$ identify the
 temperatures upto which the sample was warmed up each time after
preparing the FLL in ZFC mode in a dc bias field of 10 kOe. Thermal 
cycling was then performed by cooling down (FC) from these three
 preselected temperatures.}
\label{Fig. 6}
\end{figure}

\begin{figure}
\caption{(a),(b) Minor hysteresis curves generated by decreasing the field from 
the forward branch of the envelope loop at a fixed  temperature of 4.5 K. The 
minor curves initiated form fields lying in between $H_{pl}$ and $H_p$ do not 
merge with the
reverse leg of the envelope hysteresis loop, this gets clearly demonstrated in 
the expanded version of the data in the  
main panel of Fig. 7(b).The inset of Fig. 7(b) shows that the minor 
magnetization loops eventually merge into each other near $H_{pl}^r$ on the 
reverse leg on the envelope magnetization loop.}
\label{Fig. 7 (a),(b)}
\end{figure}

\begin{figure}
\caption{$J_c$ vs $H$ behaviour for the FLL prepared in the ZFC and FC 
modes at T = 4.5 K, shown on a log-log plot. On the log-log scale, $J_c^{ZFC}$ 
displays a linear behaviour, which accounts to a power law variation with an 
exponent of $\sim$ -1 ( i.e., $J_c^{ZFC}$ $\sim$ 1/H). Such a power 
law dependence points to a collectively pinned elastic solid 
and hence the marked region in the $(H,T)$ phase space describes the elastically 
pinned state of vortex matter. Note that the differences between the FC and ZFC
critical current density values vanish at the peak position of the PE.}
\label{Fig. 8}
\end{figure}

\begin{figure}
\caption{$\chi\prime(T)$ responses obtained for a fixed dc bias field of 
5.3 kOe in ZFC and FCW modes. 
The screening diamagnetic responses for the FLL states prepared in ZFC and FC 
modes at such low fields show very little differences. 
The inset focusses onto the two jumps across 
$T_{pl}$ and $T_p$ values, which are much smaller 
than the jumps observed for the applied dc bias field of 10 kOe.}
\label{Fig. 9}
\end{figure}

\begin{figure}
\caption{Computed values of radial length 
$R_c/a_0$ and longitudinal length $L_c/d$ 
for Ca$_3$Rh$_4$Sn$_{13}$ 
at T = 4.5 K for FLL prepared in ZFC 
as well as in FC mode. $a_0$ represents the
lattice constant ($\approx\phi_0/\sqrt{B}$)
 and d denotes the thickness of the sample 
in the direction of the field.
The two ratios for the 
ZFC mode initially increase with increasing 
H upto the onset field $H_{pl}$, where 
they start decreasing sharply with 
H. At H = $H_p$ , $R_c\sim a_0$, which 
indicates that the FLL gets completely
amorphized at the peak of the PE with 
the the total loss of radial correlation.}
\label{Fig. 10}
\end{figure}

\begin{figure}
\caption{Vortex phase diagram for Ca$_3$Rh$_4$Sn$_{13}$ depicting
($H_{pl}$,$T_{pl}$), ($H_p$, $T_p$), ($H_{irr}$, $T_{irr}$), ($H_{c2}$
, $T_c$) lines. For an explanation of nomenclature of vortex phases in different  (H, T) regions, see text.}
\label{Fig. 11}
\end{figure}

\begin{table}
\caption{Superconducting parameters of Ca$_3$Rh$_4$Sn$_{13}$ using isotropic GL theory at $T~=~4.5~K$}
\begin{tabular}{ccccccccc}
$\lambda$($\AA$)&$\kappa$&$\xi(\AA)$&$H_{c1}$(Oe)&$H_{c2}$(Oe)&$H_c$(Oe)&Gi&$j_0(
$A/m$^2$)&$j_c$(H = 0)(A/m$^2$)\\
\tableline

2270&17.5&130&110&19500&800&3~$\times$~10$^{-
7}$&1.5~$\times$~10$^{11}$&10$^7$\\

\end{tabular}

\label{table 1}
\end{table}


\begin{references}

\bibitem{r1} S. Bhattacharya and M. J. Higgins, Phys. Rev. 
Lett. {\bf70}, 2617 (1993); M. J. Higgins 
and S. Bhattacharya, Physica C {\bf257},  232 (1996) 
and references therein. 

\bibitem{r2} A. D. Huxley, C. Paulsen, O. 
Laborde, J. L. Tholence, D. Sanchez,
 A. Junod and R. Calemczuk, J. Phys. Condens. 
Matter {\bf5}, 7709 (1993); S. B. Roy and P. Chaddah,
 $ibid.$ {\bf9}, L625 (1997). 

\bibitem{r3} R. Modler, P. Gegenwart, M. Lang,
 M. Deppe, M. Weiden, T. Luhmann, C. Geibel,
 F. Steglich, C. Paulsen, J. L. Tholence, N. Sato, T. Komatsubara,
 Y. Onuki, M. Tachiki and S. Takahashi, Phys. Rev. Lett. 
{\bf76}, 1292 (1996).

\bibitem{r4} M. Tachiki, S. Takahashi,
 P. Gegenwart, M. Weiden, M. Lang, 
C. Geibel,
 F. Steglich, R. Modler,
 C. Paulsen and Y. Onuki, Z. Phys. B {\bf100}, 
369 (1996).

\bibitem{r5} N. R. Dilley, 
J. Herrmann, S. H. Han, M. B. Maple, S. Spagne,
 J. Diederichs and R. E. Sager, Physica C {\bf265}, 
150 (1996).

\bibitem{r6} S. B. Roy, Phil. Mag.
 {\bf65}, 1453 (1992); K. Yagasaki, M. Hedo and 
T. Nakama, J. Phys. Soc. Jpn. 
{\bf62}, 3825 (1993); S. B. Roy, P. Chaddah
 and S. Chaudhary, J. Phys. Condens. Matter {\bf10}, 4885 (1998). 

\bibitem{r7} P. L. Gammel,  U. Yaron,  A. P. Ramirez,  D. J. 
Bishop, A. M. Chang, R. Ruel, L. N. Pfeiffer and E. Bucher, G. D'Anna,
 D. A. Huse, K. Mortensen, M. R. Eskildsen, P. H. Kes, Phys. Rev. Lett. {\bf80}, 833 (1998).

\bibitem{r8} M. Isino, T. Kobayashi, N. Toyota, T. Fukase 
and Y. Muto, Phys. Rev. B {\bf38}, 4457 (1988).

\bibitem{r9} J. G. Park,  M. Ellerby,  K. A. McEwen and M. de 
Podesta, J. Magn. Magn. Mater. {\bf140-144}, 2057 (1995).

\bibitem{r10} U. Yaron, P. L. Gammel,
 D. A. Huse, R. N. Kleiman, C. S. Oglesby,
 E. Bucher, B. Batlogg, D. J. Bishop, K. Mortensen,
 K. Clausen, C. A. Bolle and F. de la Cruz, Phys. Rev. Lett. 
{\bf73}, 2748 (1994).

\bibitem{r11} C. A. Bolle, F. de la Cruz,  P. L. Gammel,
J. V. Waszczak and D. J. Bishop, Phys. Rev. Lett. {\bf71}, 
4039 (1993); F. Pardo, F. de la Cruz, P. L. Gammel, 
C. S. Oglesby,
 E. Bucher, B. Batlogg and D. J. Bishop,
Phys. Rev. Lett. 
{\bf78}, 4633 (1997).

\bibitem{r12} H. Sato,  Y. Akoi,  H. Sugawara and T. 
Fukahara, J. Phys. Soc. Jpn.  {\bf64}, 3175 (1995).

\bibitem{r13} C. V. Tomy,  G. Balakrishnan and D. Mck. Paul, 
Physica C {\bf280}, 1 (1997).

\bibitem{r14} K. Hirata {\rm et al}., $Advances~in~
Sperconductivity~VIII$, edited by H. Hayakawa and Y. 
Enomoto (Springer-Verlag, Berlin, Germany, 1996) p. 619.

\bibitem{r15} X.~Ling~and~J.~I.~Budnick,~$Magnetic$ $Susceptibility$ $of$ $Superconductors$ $and$ $other$ $Spin$  $Systems$, edited by R. A. Hein, T. L.  Francavilla and D. H. Liebenberg 
(Plenum Press, New York, 1991), p.377.

\bibitem{r16} G. D'Anna,  W. Benoit,  W. Sadowski and E. 
Walker, Europhys. Lett. {\bf20},  167 (1992); H. K$\ddot{\rm {u}}$pfer, 
T. Wolf, C. Lessing, A. A. Zhukov, X. Lancon, R. Meir-Hirman and W. Sch Phys. Rev. 
B {\bf58}, 2886 (1998) and references therein.

\bibitem{r17} A. B. Pippard, Philos. Mag. {\bf19},  217 
(1969); A. M. Campbell and J. E. Evetts, Adv. Phys. 
{\bf21}, 327 (1972) and references therein.


\bibitem{r18} P. Fulde and R. A. Ferrel,  Phys. Rev. 
{\bf135A}, 550 (1964).

\bibitem{r19} A. I. Larkin and Y. N. Ovchinnikov,  Sov. Phys. 
JETP {\bf20}, 762 (1965). 

\bibitem{r20} K. Gloos, R. Modler, 
H. Schimanski, C. D. Bredl, C. Geibel, 
F. Steglich, A. I. Buzdin, N. Sato and 
T. Komatsubara, Phys. Rev. Lett. {\bf70}, 501 (1993); A. Ishiguro, A. Sawada, Y. Inada, J. Kimura, M. Suzuki, N. Sato and T. Komatsubara, J. Phys. 
Soc. Jpn. {\bf64}, 378 (1995).


\bibitem{r21} G. W. Crabtree,  M. B. Maple,  W. K. Kwok,  J. 
Herrmann, J. A. Fendrich,  N. R. Dilley and  S. H. Han, Physics 
Essays {\bf9}, 628 (1996). 

\bibitem{r22} C. Tang,  X. S. Ling,  S. Bhattacharya and 
P. M. Chaikin, Europhys Lett. {\bf35}, 597 (1996) and 
references therein.

\bibitem{r23} S. S. Banerjee, N. G. Patil,
 S. Saha, S. Ramakrishnan, A. K. Grover, S. Bhattacharya,
 G. Ravikumar, P. K. Mishra, T. V. C. Rao,
V. C. Sahni, M. J. Higgins, E. Yamamoto, Y. Haga, M. Hedo, Y. Inada
 and Y. Onuki, Phys. Rev. B. 
{\bf58}, 995 (1998).

\bibitem{r24} C. V. Tomy, G. Balakrishnan and D. McK. Paul,
 Phys. Rev. B {\bf56}, 
8346 (1997).

\bibitem{r25} G. Ravikumar, V. C. Sahni, P. K. Mishra, 
T. V. C. Rao, S. S. Banerjee, A. K. Grover,
 S. Ramakrishnan, S. Bhattacharya, 
M. J. Higgins, E. Yamamoto,
 Y. Haga, M. Hedo, Y. Inada and Y. Onuki, Phys. Rev. B 
{\bf57}, R11069 (1998).

\bibitem{r26} W. Henderson, E. Y. Anderi, M. J. Higgins
 and S. Bhattacharya, Phys. Rev. Lett. 
{\bf77}, 2077 (1996); {\bf80}, 381 (1998).

\bibitem{r27} S. Ramakrishnan, S. Sundaram, R. S. Pandit and G. Chandra, 
J. Phys. E {\bf18}, 650 (1985).

\bibitem{r28} G. Ravikumar, T. V. C. Rao, P. K. Mishra, 
V. C. Sahni, S. S. Banerjee, A. K. Grover,
 S. Ramakrishnan, S. Bhattacharya, 
M. J. Higgins, E. Yamamoto,
 Y. Haga, M. Hedo, Y. Inada and Y. Onuki, Physica C 
{\bf276}, 9 (1997); {\bf298}, 122 (1998).

\bibitem{r29} C. P. Bean,  Rev. Mod. Phys. {\bf36}, 31 
(1964).

\bibitem{r30} S. Kokkaliaris, D. A. J. de Groot, 
S. N. Gordeev, A. A. Zhukov, R. Gagnon and
 L. Taillefer, Phys. Rev. Lett. {\bf82}, 5116 (1999).

\bibitem{r31} A. Huxley, R. Cubitt, D. McK. Paul, E. Forgan,
 M. Nutley, H. Mook, M. Yethiraj, P. Lejay, D. Caplan and 
J. M. Penisson, Physica B
 {\bf223~$\and$~224} 169 (1996). 

\bibitem{r32} J. A. Mydosh, $Spin ~Glasses: ~An~ 
Experimental ~Introduction$, Taylor and Francis,  London, U.K., 
1993.

 \bibitem{r33} P. Chaddah, in 
$Studies~in~High~Temperature~Superconductors$,
edited by A. V. Narlikar, Nova Science Inc., Comack, NY, USA, 1995, Vol.  
{\bf14}, pp 245-274.


\bibitem{r34} A. K. Grover, in 
$Studies~in~High~Temperature~Superconductors$,
edited by A. V. Narlikar, Nova Science Inc., Comack, NY, USA, 1995, Vol. 
{\bf14}, pp 185-244.

\bibitem{r35} S. S. Banerjee, N. G. Patil,
 S. Ramakrishnan, A. K. Grover, S. Bhattacharya,
 G. Ravikumar, P. K. Mishra, T. V. C. Rao,
V. C. Sahni and M. J. Higgins, Appl. Phys. Lett. {\bf74}, 126 (1999).

\bibitem{r36} S. B. Roy, P. Chaddah, S. Chaudhary and L. F. Cohen,
 Proceedings of the 41st Annual DAE Solid State Physics Symposium,
 Universities Press, Hyderabad (India) {\bf41}, 367 (1998). 

\bibitem{r37} S. Sarkar {\rm et al}., unpublished.


\bibitem{r38} G. Blatter, M. V. Feigel'man,
 V. B. Geshkenbein, A. I. Larkin and V. M. Vinokur, Rev. Mod. Phys. {\bf66}, 1125 (1994).

\bibitem{r39} A. I. Larkin and Y. N. Ovchinnikov, 
Sov. Phys. JETP {\bf38}, 854 (1974); A. I. Larkin, J. Low. Temp. Phys. {\bf34}, 409 (1979).

\bibitem{r40} R. Wordenweber,  P. H. Kes and C. C. Tsuei,  Phys. 
Rev. B {\bf33}, 3172 (1986); L. A. Angurel, F. Amin, M. Polichetti, 
J. Aarts and P. H. Kes, Phys. Rev. B 
{\bf56}, 3425 (1997).


\bibitem{r41} T. Giamarchi and P. Le. Doussal, 
Phys. Rev. 
Lett. {\bf72}, 1530  (1994).

\bibitem{r42} T. Giamarchi and P. Le. Doussal, Phys. Rev. B 
{\bf52},  242 (1995).

\bibitem{r43} M. Gingras and D. A. Huse,  Phys. Rev. B 
{\bf53},  15193 (1996). 

\bibitem{r44} A. Durate, E. F. Righi, C. A. Bolle, F. de la Cruz,
P. L. Gammel, C. S. Oglesby, E. Bucher, B. Batlogg and D. J. Bishop, 
Phys. Rev. B {\bf53},
 11336, (1996).

\bibitem{r45} S. S. Banerjee, N. G. Patil,
 S. Ramakrishnan, A. K. Grover, S. Bhattacharya,
 G. Ravikumar, P. K. Mishra, T. V. C. Rao,
V. C. Sahni, M. J. Higgins, C. V. Tomy, G. Balakrishnan 
and D. McK. Paul, Phys. Rev. B {\bf59}, 6043 (1999).

\bibitem{r46} S. Bhattacharya and M. J. Higgins,  Phys. Rev. B 
{\bf43}, 10005 (1995).

\bibitem{r47} S. Bhattacharya and M. J. Higgins,  Phys. Rev. B 
{\bf52}, 64 (1995).

\bibitem{r48} S. Ryu,  M. Hellerquist,  S. Doniach,  A. 
Kapitulnik and D. Stroud, Phys. Rev. Lett. {\bf77}, 5114 
(1997). 

\bibitem{r49} M. C. Faleski,  M. C. Marchetti and A.A. 
Middleton, Phys. Rev. B {\bf54}, 12427 (1996).

\bibitem{r50} K. A. M$\ddot{\rm {u}}$ller, M. Takashige and J. G. Bednorz, 
Phys. Rev. Lett. {\bf58}, 1143 (1987).

\bibitem{r51} P. Chaddah and G. Ravikumar, Pramana - J. Phys. {\bf31}, L141 (1988). 


\bibitem{r52} E. H. Brandt, J. Low. Temp. Phys. {\bf26}, 
709 (1977); {\bf26}, 735 (1977); {\bf28}, 263 (1977); 
{\bf28}, 291 (1977).

\bibitem{r53} V. M. Vinokur, P. H. Kes and A. E. Koshelev, Physica C
{\bf248}, 179 (1995).

\end{references}
\end{document}